\documentclass[%
 aip,
 amsmath,amssymb,
 preprint, floatfix,
]{revtex4-1}

\usepackage{graphicx}% Include figure files
\usepackage{dcolumn}% Align table columns on decimal point
\usepackage{bm}% bold math
\usepackage[mathlines]{lineno}% Enable numbering of text and display math
\usepackage[utf8]{inputenc}
\usepackage[T1]{fontenc}
\usepackage{mathptmx}
\usepackage{etoolbox}
\usepackage{textgreek}
\usepackage{xcolor}

\makeatletter
\def\@email#1#2{%
 \endgroup
 \patchcmd{\titleblock@produce}
 {\frontmatter@RRAPformat}
 {\frontmatter@RRAPformat{\produce@RRAP{*#1\href{mailto:#2}{#2}}}\frontmatter@RRAPformat}
 {}{}
}%
\makeatother
\begin{document}

\preprint{AIP/123-QED}

\title{High-resolution cryoEM structure determination of soluble proteins after soft-landing electrospray ion beam deposition.}
% Force line breaks with \\
\author{Lukas Eriksson}
\affiliation{Kavli Institute for Nanoscience Discovery, Dorothy Crowfoot Hodgkin Building, University of Oxford, South Parks Road, Oxford OX1 3QU, UK}
\affiliation{Department of Chemistry, University of Oxford, Mansfield Road, Oxford OX1 3TA, UK}

\author{Tim K. Esser}
 \altaffiliation[Currently at ]{Thermo Fisher Scientific, De Schakel 2, 5651GH Eindhoven, Netherlands}
\affiliation{Kavli Institute for Nanoscience Discovery, Dorothy Crowfoot Hodgkin Building, University of Oxford, South Parks Road, Oxford OX1 3QU, UK}

\author{Marko Grabarics}
\affiliation{Kavli Institute for Nanoscience Discovery, Dorothy Crowfoot Hodgkin Building, University of Oxford, South Parks Road, Oxford OX1 3QU, UK}
\affiliation{Department of Chemistry, University of Oxford, Mansfield Road, Oxford OX1 3TA, UK}

\author{Laurence T. Seeley}
\affiliation{Kavli Institute for Nanoscience Discovery, Dorothy Crowfoot Hodgkin Building, University of Oxford, South Parks Road, Oxford OX1 3QU, UK}
\affiliation{Department of Chemistry, University of Oxford, Mansfield Road, Oxford OX1 3TA, UK}
\affiliation{Department of Biochemistry, University of Oxford, South Parks Road, Oxford OX1 3QU, UK}

\author{Simon B. Knoblauch}
\affiliation{Kavli Institute for Nanoscience Discovery, Dorothy Crowfoot Hodgkin Building, University of Oxford, South Parks Road, Oxford OX1 3QU, UK}
\affiliation{Department of Chemistry, University of Oxford, Mansfield Road, Oxford OX1 3TA, UK}

\author{Jingjin Fan}
\affiliation{Kavli Institute for Nanoscience Discovery, Dorothy Crowfoot Hodgkin Building, University of Oxford, South Parks Road, Oxford OX1 3QU, UK}
\affiliation{Department of Chemistry, University of Oxford, Mansfield Road, Oxford OX1 3TA, UK}

\author{Joseph Gault}
 \altaffiliation[Currently at ]{AstraZenica, Discovery Sciences, Biopharma R\&D, Cambridge CB2 0AA, UK}
\affiliation{Department of Chemistry, University of Oxford, Mansfield Road, Oxford OX1 3TA, UK}

\author{Paul Fremdling}
 \altaffiliation[Currently at ]{REELEMENTS GmbH, Wilhelm-Eichler-Straße 34, 01445 Radebeul, Germany}
\affiliation{Department of Chemistry, University of Oxford, Mansfield Road, Oxford OX1 3TA, UK}

\author{Thomas Reynolds}
\affiliation{Department of Biology, University of Oxford, South Parks Road, Oxford OX1 3RB, UK}

\author{Justin L.P. Benesch}
\affiliation{Kavli Institute for Nanoscience Discovery, Dorothy Crowfoot Hodgkin Building, University of Oxford, South Parks Road, Oxford OX1 3QU, UK}
\affiliation{Department of Chemistry, University of Oxford, Mansfield Road, Oxford OX1 3TA, UK}

\author{Carol V. Robinson}
\affiliation{Kavli Institute for Nanoscience Discovery, Dorothy Crowfoot Hodgkin Building, University of Oxford, South Parks Road, Oxford OX1 3QU, UK}
\affiliation{Department of Chemistry, University of Oxford, Mansfield Road, Oxford OX1 3TA, UK}

\author{Jani R. Bolla}
\affiliation{Department of Biochemistry, University of Oxford, South Parks Road, Oxford OX1 3QU, UK}
\affiliation{Department of Biology, University of Oxford, South Parks Road, Oxford OX1 3RB, UK}

\author{Lindsay Baker}
\affiliation{Kavli Institute for Nanoscience Discovery, Dorothy Crowfoot Hodgkin Building, University of Oxford, South Parks Road, Oxford OX1 3QU, UK}
\affiliation{Department of Biochemistry, University of Oxford, South Parks Road, Oxford OX1 3QU, UK}

\author{Stephan Rauschenbach}
 \email{stephan.rauschenbach@chem.ox.ac.uk}
\affiliation{Kavli Institute for Nanoscience Discovery, Dorothy Crowfoot Hodgkin Building, University of Oxford, South Parks Road, Oxford OX1 3QU, UK}
\affiliation{Department of Chemistry, University of Oxford, Mansfield Road, Oxford OX1 3TA, UK}

\date{\today}% It is always \today, today,
 % but any date may be explicitly specified

\begin{abstract}
A protein's structure in the gas phase underpins the interpretation of native mass spectrometry. Yet how fold and conformation respond to dehydration has never been resolved at the residue level, due to the unavailability of a method to image gas-phase proteins at near-atomic resolution. Here, we determine near-atomic-resolution cryoEM structures (2.5--4.8\,\AA) of four soluble protein complexes ($\beta$-Galactosidase, GDH, RuBisCo, and GroEL) prepared by soft-landing electrospray ion beam deposition (ESIBD) and show the retention of secondary and tertiary structure. Comparison with the corresponding solution structures reveals dehydration-induced structural change is governed by local solvent exposure: interior residues retain high-resolution density while solvent-exposed regions are likely to rearrange. Coherent rearrangements preserve secondary and tertiary structure, incoherent changes manifest as local loss of resolution. Dedicated instrumentation provided the required control over deposition energy, sample environment, and growth of thin vitreous ice films embedding the landed proteins. ESIBD+cryoEM thereby links the chemical selectivity of native mass spectrometry directly to near-atomic structural resolution.
\end{abstract}

\maketitle

\section*{INTRODUCTION}
Establishing the relationship between a protein's precise chemical identity, its structure, and its interactions is central to many questions in molecular biology, (patho)physiology, and drug discovery.\cite{IL2016, HB2025} X-ray crystallography and cryo-electron microscopy (cryoEM) are widely employed for structural characterisation, and mass spectrometry (MS) for chemical characterisation. Native MS, based on the gentle, intact electrospray ionisation (ESI) of protein complexes with their ligands bound, excels at determining mass, stoichiometry, binding, interaction strength, and sample heterogeneity.\cite{JG2020}

However, native MS yields no direct information about the three-dimensional structure of the gas-phase protein ion, leaving the structural consequences of the transfer from solution to vacuum uncharacterised. Ion mobility spectrometry (IMS) indicates that gas-phase proteins adopt more compact configurations than predicted from their solution structures,\cite{MB2010, KJ2021, CH2011a, ED2009, EM2015} but the microscopic basis of this compaction---the identity of the rearranging residues, the pathways of rearrangement, and the forces driving it---has remained inaccessible.\cite{EC2023, SC2018}

Direct observation of gas-phase protein structure at the residue level requires a method capable of preparing and imaging dehydrated protein complexes at near-atomic resolution. Soft-landing electrospray ion beam deposition (ESIBD) combined with cryoEM provides a route to such observation. In ESIBD, intact molecular gas-phase ions, generated by ESI, are transferred into vacuum, purified by mass filtering, and deposited onto surfaces under controlled conditions.\cite{SR2016} This precisely controlled process ensures chemical purity, homogeneity of the prepared samples, and gentle delivery to the surface, as demonstrated extensively in sample fabrication for single molecule imaging by scanning probe microscopy (SPM).\cite{XW2020, KA2023} CryoEM, a primary tool for protein structure determination,\cite{YC2015, EC2020} achieves near-atomic resolution by averaging over many particles from transmission electron micrographs of purified, homogeneous samples. In the ESIBD-cryoEM combination, the chemical selectivity of native MS is linked directly to high-resolution protein structure determination.

Initial proof-of-concept experiments established the feasibility of this coupling at low resolution combining room-temperature landing with cryoEM\cite{TE2022a, PF2022, TE2022} or negative-stain EM imaging.\cite{MW2022} For a single protein, $\beta$-Galactosidase, we subsequently demonstrated structure determination at 2.6\,\AA~using cryogenic landing, ice embedding, and cryoEM imaging.\cite{TE2024} Secondary and tertiary structure were largely retained, although a slight compaction of the protein and a reduction of resolution at its surface were observed. However, the generality of this approach across proteins of varying size and oligomeric state remains unclear, and the principles governing gas-phase restructuring are not well understood.

Here we report near-atomic-resolution cryoEM structures of four soluble protein complexes prepared by ESIBD---$\beta$-Galactosidase, glutamate dehydrogenase (GDH), RuBisCo, and GroEL---spanning a broad range of molecular masses and oligomeric stoichiometries. Comparison of these gas-phase-derived structures with their solution counterparts reveals that dehydration-induced structural change is governed by local solvent exposure. Generally, residues located in the protein interior, shielded from solvent in solution, retain high-resolution density in the resulting maps, whereas surface-exposed regions rearrange. The rearrangement proceeds either along a coherent pathway, in which all proteins in the population rearrange in the same way causing no loss in resolution, or via incoherent pathways, where a protein or region rearranges along multiple trajectories, causing it to appear as blurred density. We observe coherent rearrangement in the inter-ring twist and axial compaction of GroEL and in the crevice closure of $\beta$-Galactosidase. In both cases secondary and tertiary structure are preserved and domain-level movement is resolved. Incoherent rearrangement is observed in the collapse of the GDH antenna helices and in surface sidechains of RuBisCo, where it manifests as local resolution loss. A solvent exposure score computed from the solution structure predicts both outcomes and provides a microscopic account of the gas-phase compaction previously inferred from IMS measurements. Understanding these dehydration-induced changes opens a route towards their targeted reversal, for instance by laser flash melting of the embedding ice.\cite{SB2025}

Achieving these results consistently requires precise control over the protein and its environment throughout the ESIBD process. While several (electro-)spray methods for cryoEM preparation have been demonstrated,\cite{ZY2024, AG2025} the ESIBD approach goes well beyond ambient-pressure spray deposition without desolvation or mass selection. It includes gentle native ESI, vacuum transfer, thermalisation and transport of the gas phase ion, mass analysis and selection, cryogenic, contamination-free, and energy-controlled deposition, and the reproducible growth of thin, vitreous ice films embedding the landed proteins. We describe the instrumentation and workflow developed to meet these requirements, establish the conditions under which samples suitable for high-resolution cryoEM are obtained, and discuss the implications for the interpretation of native MS data and for structural biology based on mass-selected vacuum deposition.

\section*{RESULTS}

\subsection*{Instrumentation and Workflow}

\begin{figure}%[h]
 \centering
 \includegraphics[width=\linewidth]{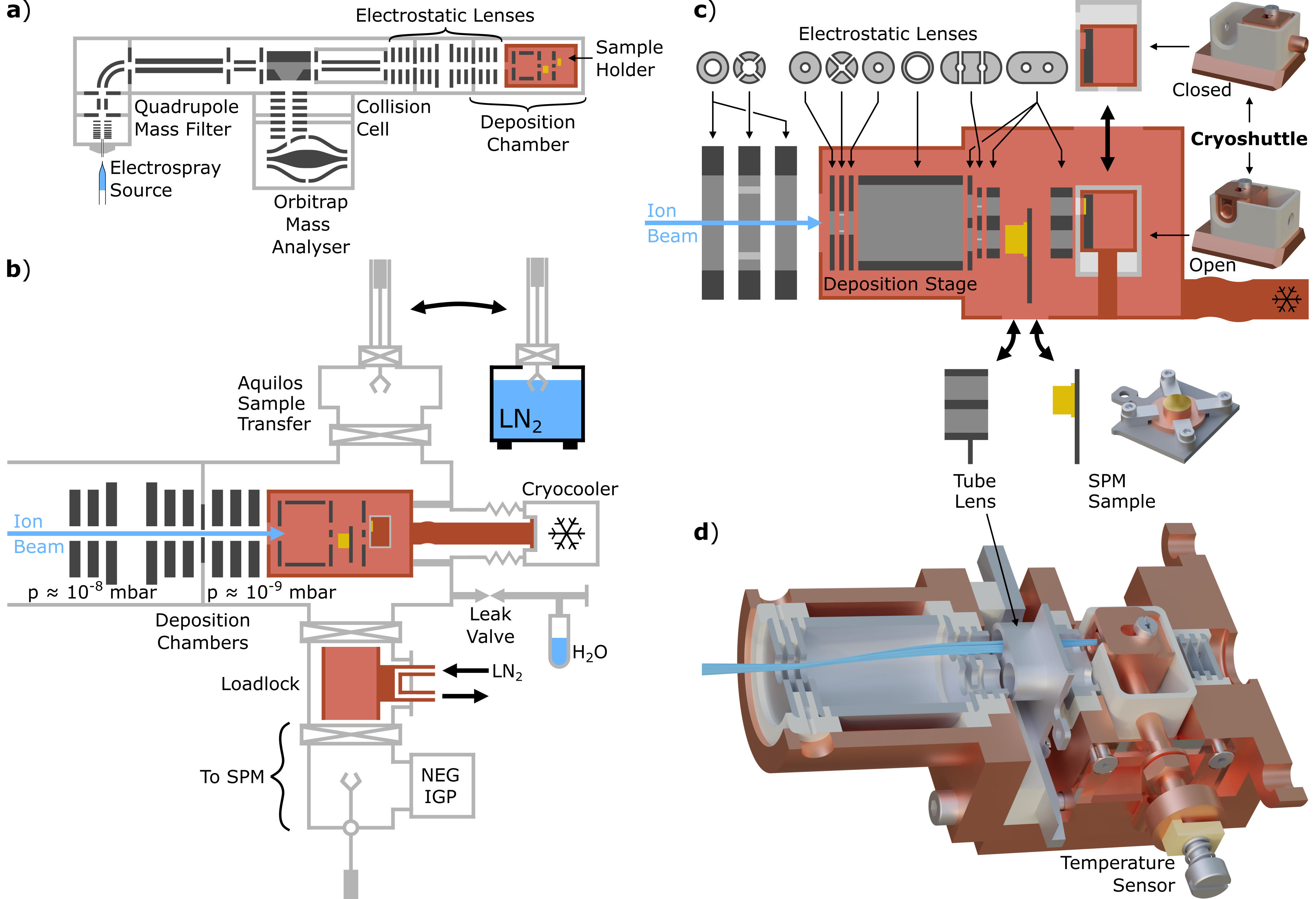}
 \caption{\textbf{Instrumentation Overview.} \textbf{a)} Schematic of the deposition instrument based on a modified Orbitrap UHMR. \textbf{b)} Schematic of the deposition chamber containing the cryogenic sample stage and sample transfer systems for cryoEM and SPM samples. \textbf{c)} Cross section schematic of the cryogenic deposition stage. Cryoshuttle shown in both states: deposition (open) and transfer (closed). \textbf{d)} Section-view rendering of the deposition stage with simulated ion trajectories (blue) for deposition onto a cryoEM grid. Tube lens inserted into SPM sample holder position.}
 \label{fig_overview}
\end{figure}

For ESIBD experiments, we use a modified Thermo Scientific Q Exactive UHMR instrument,\cite{PF2022} extended by two differential pumping stages containing electrostatic lenses and deposition instrumentation reaching ultrahigh vacuum (UHV, $p<10^{-9}$ mbar, Fig.\,\ref{fig_overview}a) at the sample position. A deposition extension like this can be fitted, in principle, to any typical native ESI-MS instrumentation, and ESIBD workflows can be applied for SPM\cite{AW2022, SR2016, GJ2011}, for cryoEM\cite{NV2018, MW2022}, or for sample fabrication of other methods.\cite{CH2011, HS2021, VF1977} 

Molecular ions are generated by native ESI. After m/z-filtering and characterisation by MS, the ion beam exits the collision cell through an aperture with well-defined chemical composition, energy, and spatial distribution. Electrostatic (DC) ion optics (Fig.\,\ref{fig_overview}) steer and focus the beam onto cryoEM grids held within a cryoshuttle (Fig.\,\ref{fig_overview}c) or onto single crystal surfaces for SPM imaging (see SI for details on SPM capabilities), where the landing energy is defined by the bias of the sample. Beam currents at apertures and at the sample are measured by picoammeters, aiding alignment and focusing of the ion beam and allowing precise quantification of the deposited amount of material.

For structure determination by cryoEM, it is crucial that the sample is kept at a consistent, cryogenic temperature to avoid thermal activation, which would introduce structural heterogeneity and degrade resolution.\cite{MW2023,TE2024} The deposition stage is cooled by a reverse Stirling cryocooler (Fig.\,\ref{fig_overview}b) to a temperature between 60 K and 350 K. To minimise sample contamination by adsorption of residual gases, the deposition stage is designed such that sample surfaces do not have direct line of sight to any room temperature surfaces, and UHV is achieved in the deposition chamber (Fig.\,\ref{fig_overview}b). After deposition, the landed protein molecules can be embedded in ice by leaking water vapour into the deposition chamber at a defined partial pressure.

To protect cryoEM grids from contamination during transfer, a cryoshuttle is employed (compatible with Thermo Scientific's Aquilos transfer hardware). During transfer, grids are protected by PEEK shutters which close as the shuttle is removed from the deposition position (Fig.\,\ref{fig_overview}c, suppl.\,movie\,S1). The shuttle is transferred through static vacuum and clean nitrogen gas into liquid nitrogen (Fig.\,\ref{fig_overview}b).\cite{TE2024} Throughout transfer, which takes less than a minute, the grid remains below the glass transition temperature of ice ($\sim$136\,K),\cite{ST2021, RS2000} preventing devitrification and contamination. 

\subsection*{Ice growth on TEM grids}

In the conventional plunge freezing sample preparation for high-resolution cryoEM, particles are embedded in a vitreous ice layer that is sufficiently thin (10--50\,nm) and homogeneous in phase and morphology.\cite{LP2016, GW2021}

\begin{figure}%[h]
\centering
\includegraphics[width=\linewidth]{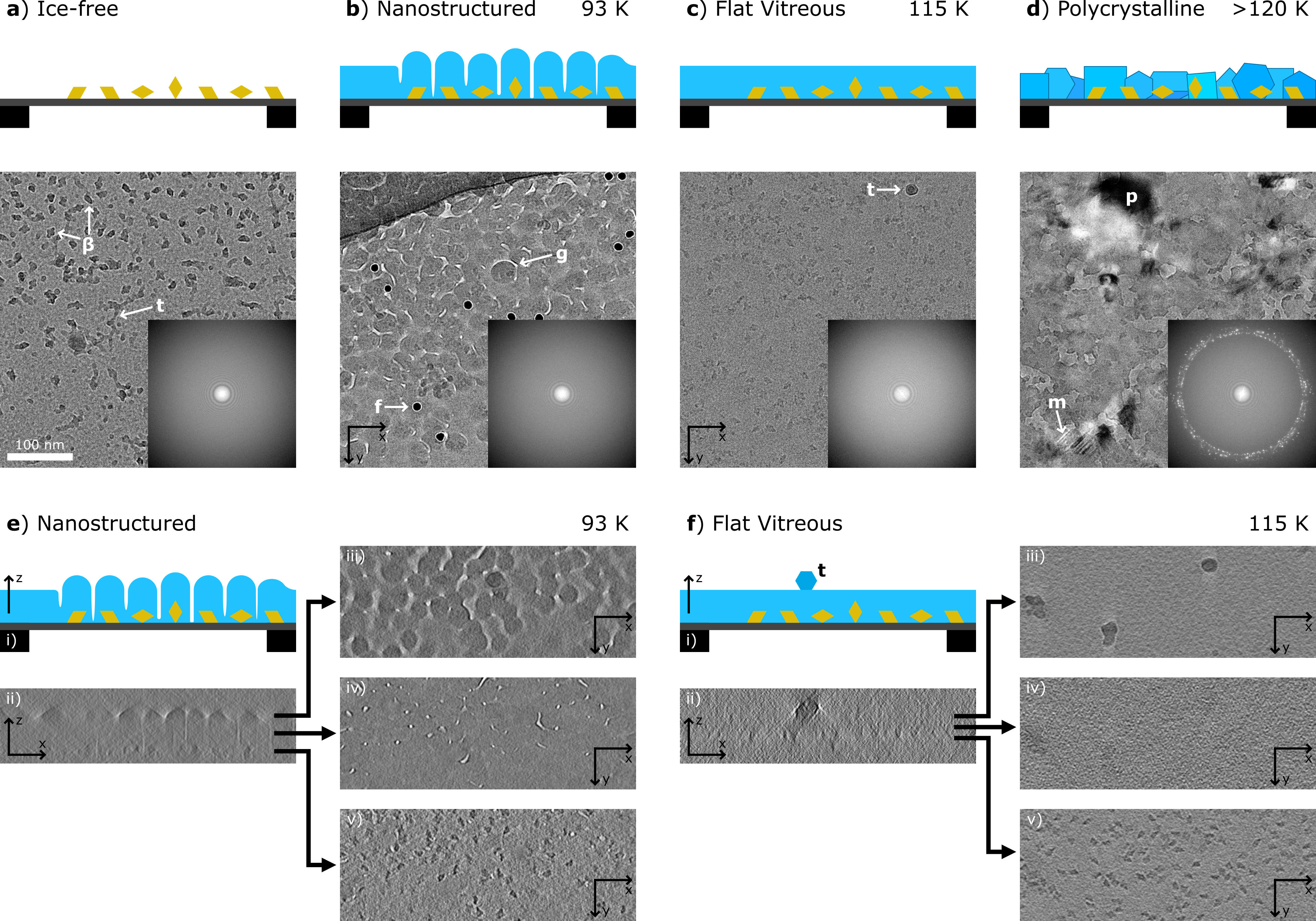}
\caption{\textbf{Control of ice phase and morphology.} \textbf{a-d}) Side view schematics, cryoEM micrographs, and inset power spectra of $\beta$-Galactosidase deposited on amorphous carbon films \textbf{a}) without ice deposition. \textbf{b}) Ice grown at 93 K. \textbf{c}) Ice grown at 115 K. \textbf{d}) Ice grown above 120 K. All ice was grown for 3 minutes at $5\times10^{-5}\,\mathrm{mbar}$ and all micrographs are at the same scale. $\beta$-galactosidase particles ($\beta$), transfer ice contamination (t), gold fiducials (f), gaps (g), and features indicative of crystalline ice (p, m) are labelled. \textbf{e}) Schematic (\textbf{i}) and tomogram slices (\textbf{ii-v}) of nanostructured ice, grown at 93 K. Tomogram xz-slice shows columns of ice with rounded tops. Proteins can be seen towards the bottom as darker patches and in the xy-plane slice (v). The gaps between nanostructures appear as bright, vertical lines (in xz slice), and higher xy-slices show nanostructured ice and the rounded tops. \textbf{f}) Schematic (i) and tomographic slices (ii) xz, (iii-v) xz at various heights show evidence of a flat, vitreous ice film grown at 115 K embedding the proteins.
}\label{fig_ice}
\end{figure}

Proteins prepared by ESIBD adsorb on the surface of an amorphous carbon support film,\cite{TE2022} in vacuum. In the absence of ice, they show very high contrast in micrographs (Fig.\,\ref{fig_ice}a), but an unresolved shell at the protein surface prevents high-resolution structural analysis.\cite{TE2022} Upon embedding the proteins in an ice film grown in the deposition chamber, such a shell is not observed which allows the structure to be determined at high resolution throughout.\cite{TE2024}

The optimal ice film is thin, flat, homogeneous, and vitreous (low-density amorphous) so that the embedded proteins are not obscured by contrast from crystalline grains or surface roughness.\cite{JD1988, LP2016} We grow ice films by condensing water vapour at $5\!\times\!10^{-5}\,\mathrm{mbar}$ on a TEM grid surface at well-defined cryogenic temperatures (see Methods). After ice growth, the stage is cooled to 93 K before the shuttle is transferred into liquid nitrogen.\cite{TE2024}

To explore and optimise the control of ice growth, $\beta$-Galactosidase was deposited onto 2\,nm amorphous carbon TEM grids. Ice was grown at various temperatures, and the samples were imaged by cryoEM, including tilt series for tomographic reconstruction. Figure \ref{fig_ice}a-d shows schematic cross sections, representative micrographs, and inset power spectra for four conditions: no ice growth, ice grown at 93\,K, 115\,K, and above 120\,K.

Without water-vapour exposure (Fig.\,\ref{fig_ice}a), the $\beta$-Galactosidase particles ($\beta$) appear at high contrast against the amorphous carbon background. Small ice particles (t) reflect residual contamination during sample transfer.

Ice grown at 93\,K (Fig.\,\ref{fig_ice}b) reduces particle contrast and produces rounded features of $\sim$30\,nm diameter surrounded by bright, narrow edges (g, 5--50\,nm), with proteins embedded in the ice columns. Dark spots (f) are 10\,nm gold fiducials added for tomogram reconstruction. The power spectrum confirms the ice is amorphous (no diffraction features). Tomography (Fig.\,\ref{fig_ice}e) reveals the ice grows as nanostructured columns above the proteins with rounded tops, separated by thin, crevasse-like voids consistent with self-shadowed growth.\cite{AD1977}

Ice grown at 115\,K (Fig.\,\ref{fig_ice}c) yields the ice film of desired properties: $\beta$-Galactosidase particles are clearly resolved at moderate contrast, with no crystalline features apart from a single transfer-contamination crystal (t). The ice layer is vitreous, $\sim$20\,nm thick, and homogeneous as confirmed by cryoET (Fig.\,\ref{fig_ice}f).

Ice grown above 120\,K (Fig.\,\ref{fig_ice}d) is polycrystalline, identifiable by bright and dark patches (p) and Moiré fringes (m).\cite{JD1988, YC2015, ML2024, HS2023} The power spectrum shows diffraction peaks consistent with cubic crystalline domains. Protein particles are visible but largely obscured by the polycrystalline ice.

Of the four conditions, only 115\,K ice yields samples suitable for single-particle analysis: nanostructured 93\,K ice and polycrystalline $> 120\,K$ ice both introduce contrast features that obscure the proteins or hamper alignment, while ice-free samples lack the surface-shell-removing effect of vitreous embedding.\cite{TE2024}

Precise control of stage temperature, water partial pressure, and growth time, combined with the optimised transfer into liquid nitrogen, yields consistent vitreous embedding suitable for high-resolution cryoEM.

\subsection*{ESIBD+cryoEM workflow for soluble proteins}

\begin{figure}%[h]
\centering
\includegraphics[width=\linewidth]{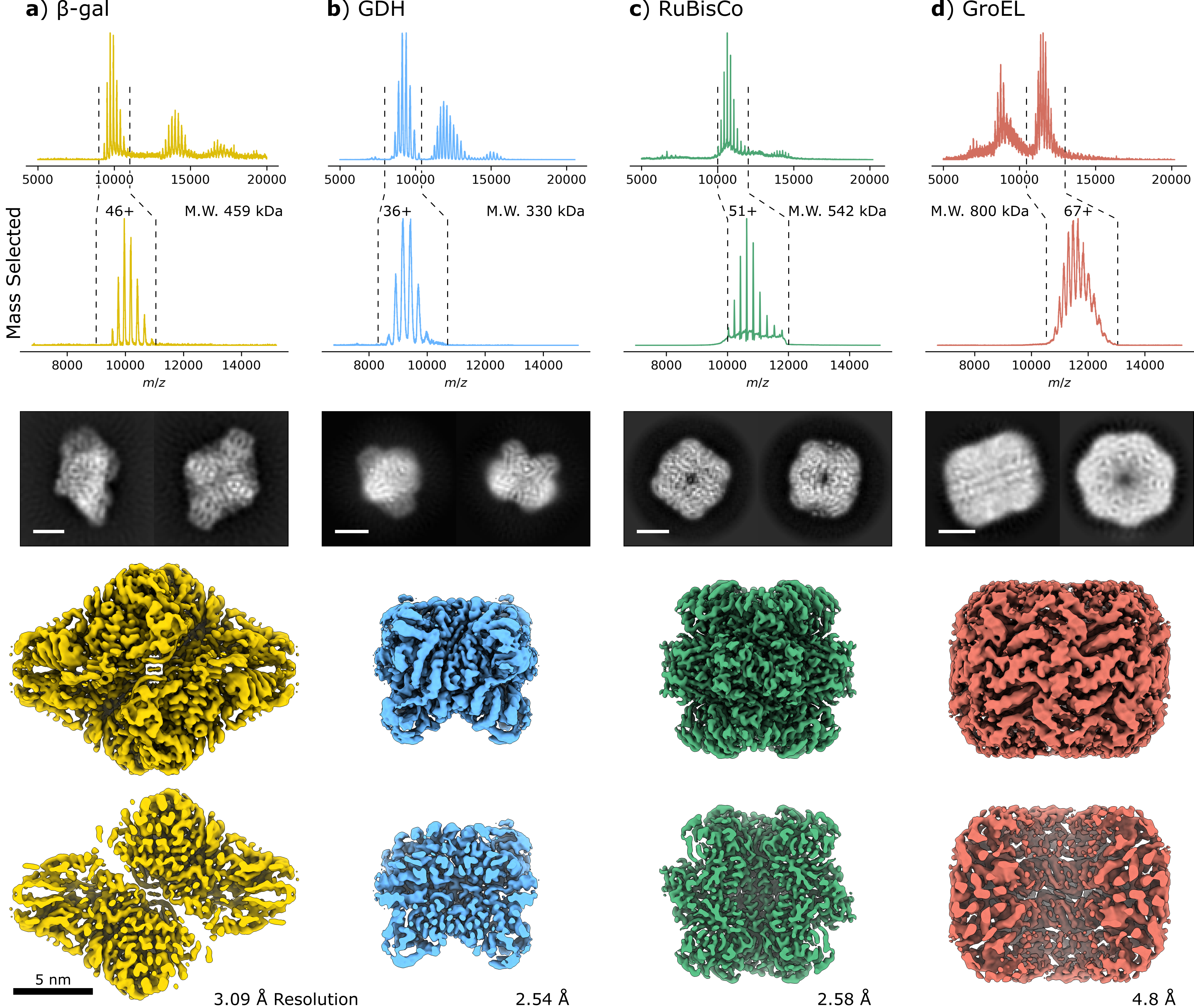}
\caption{\textbf{Mass spectra, 2D classes, and cryoEM maps of ESIBD-prepared soluble proteins.} Native MS shown before and after m/z selection (filter region indicated by dashed lines), representative 2D classes (scale bars 5 nm), and cryoEM density maps for $\beta$-Galactosidase (\textbf{a}), GDH (\textbf{b}), RuBisCo (\textbf{c}), and GroEL (\textbf{d}). Resolution values given at FSC 0.143 cutoff.}\label{fig_em}
\end{figure}

During deposition, the ions are guided through the collision cell by a gentle voltage gradient, where many low-energy ion-nitrogen collisions thermalise the ions, effectively narrowing the energy distribution of the ion beam to approximately 1 eV per charge at FWHM,\cite{PF2023} which represents the lower limit achievable for the surface-molecule collision.\cite{PF2022} After thermalisation, the ion beam enters the high vacuum part of the ESIBD instrument (see Fig. \ref{fig_overview}), where no more collisions with gas molecules occur, which means that from this point on the beam energy is defined and will be conserved until deposition. Protein ion beams are typically centred in energy around -7 eV per charge, with values given with respect to ground (this means the ion optics are biased more negatively to allow transmission). The ions are then directed to the grid for deposition which is held at a temperature of 115 K. For very gentle landing with collision energy of less than 2 eV per charge, the voltage of the grid is typically set to -8.5 V. Typically, currents in the range of 20-50 pA at the sample are achieved after mass selection. The quantity deposited is monitored by integrating the deposition current over time; the deposited charge in the following is given in pAh (picoampere-hours).\cite{SR2016}

The tight energy distribution allows for effective focusing of the ion beam onto a small region of the surface, allowing for fast depositions resulting in reasonable particle density. We landed 11 pAh of $\beta$-Galactosidase, 15 pAh of GDH and GroEL, and 10 pAh of RuBisCo, or approximately 5 to 10 billion molecules in 15-45 minutes, resulting in roughly 4 mm$^2$ of the 7.8 mm$^2$ grid having optimal particle density for data collection (see suppl.\,eq.\,1,2). An optimal density maximises the number of particles per micrograph without particles being so close together that they hamper alignment and particle picking. Ideally, particles will be spaced approximately one to two particle lengths away from all nearest neighbours, or 10 to 25\% of monolayer coverage.

After protein deposition, ice was grown at 115 K and samples were transferred to storage in liquid nitrogen as described above. From here, the samples are handled identically as samples prepared by plunge freezing.

\subsection*{Structural analysis of ESIBD-prepared proteins}

\begin{figure}%[h]
\centering
\includegraphics[width=\linewidth]{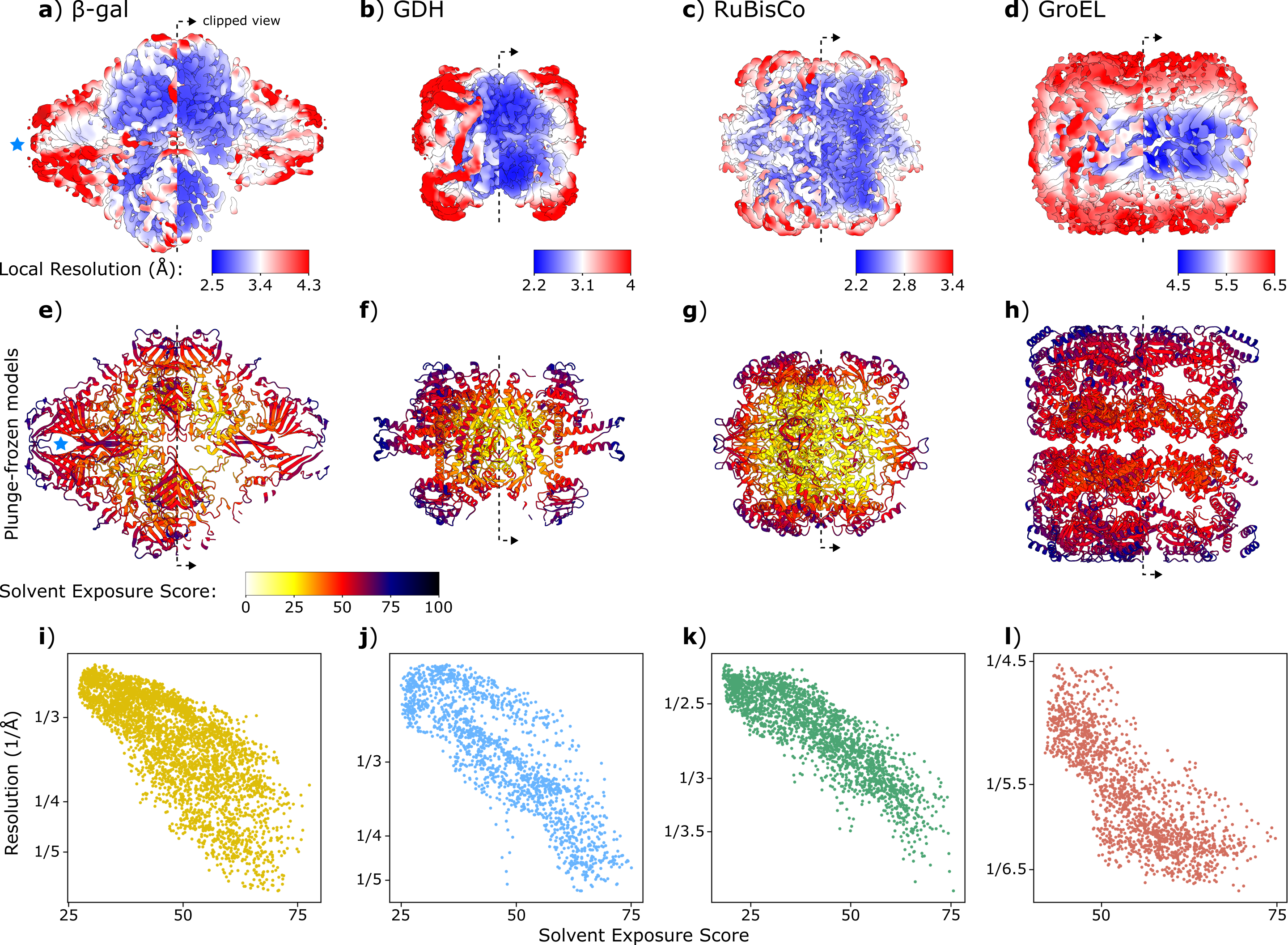}
\caption{\textbf{Local resolution trends in ESIBD-prepared proteins.} \textbf{a-d}) Consensus cryoEM maps of proteins prepared with ESIBD, coloured by local resolution (FSC cutoff 0.5). \textbf{e-h}) Solution phase models of proteins coloured by solvent exposure score. \textbf{i-l}) Plots of solvent exposure score (from solution phase model) versus local resolution for all backbone atoms modelled in ESIBD-prepared proteins. Individual maps, models, and plots for $\beta$-Galactosidase, GDH, RuBisCo, and GroEL. A blue star marks the location of a crevice which collapses in the gas phase.}\label{fig_locres}
\end{figure}

We imaged the samples and processed the resulting data with standard workflows (see methods/SI).\cite{AP2017, JZ2018} All proteins studied here, which vary in size, expression organism, and oligomeric stoichiometry, show 2D classes with characteristic features of $\alpha$-helices and $\beta$-sheets. The reconstructed consensus 3D maps (Fig.\,\ref{fig_em}), with symmetry (D2, D3, D4, and D7, respectively) imposed, are high-resolution (all below 5\,\AA) and atomic models were built into all four maps.

Comparison with the corresponding plunge-frozen reference structures reveals three consistent features. First, secondary structure motifs present in solution are largely reproduced in the ESIBD-prepared maps. Second, local resolution varies systematically with position: it is highest in the protein interior and lower at protein surfaces, with significant variation across the surface itself. Third, domain-level reorientations occur with respect to the solution structure, while the internal structure of each domain is largely preserved.

To quantify the relationship between solvent exposure and local resolution, we introduce a solvent exposure score computed from plunge-frozen or crystal structures (see Methods). For each atom, the score estimates the share of non-covalent interactions originating from surrounding solvent rather than from other protein residues, with higher values indicating greater solvent exposure.

Figure\,\ref{fig_locres}a-d shows the consensus cryoEM maps coloured by local resolution; Fig.\,\ref{fig_locres}e-h shows the corresponding plunge-frozen reference structures coloured by solvent exposure score. For $\beta$-Galactosidase, GDH, and RuBisCo, resolution of the core regions ranges from 2.2--3.5\,\AA, while GroEL is lower resolution, 4.5--6.5\,\AA. The interior of each protein, where atoms are surrounded mostly by other protein residues, yields a low solvent exposure score and high local resolution. Surface atoms, particularly those at exposed or protruding sites, yield high solvent exposure scores and lower local resolution. This trend, previously noted for $\beta$-Galactosidase alone,\cite{TE2024} holds across all four proteins studied here, with the solvent exposure score correlating quantitatively with local resolution (Fig.\,\ref{fig_locres}i-l).

The solvent exposure score also captures the variation across the protein surface itself. The inside of the cavity of RuBisCo exhibits a very low solvent exposure score and is reproduced at high resolution (<2.5 \AA, Fig. 4c). Conversely, the protruding antenna helices of GDH carry very high solvent exposure scores (Fig. 4b), and the corresponding density is resolved only at low map threshold (Fig. 5c).

Beyond per-residue resolution, the solvent exposure score also predicts where domain-level reorientation occurs. Morphological features such as surface crevasses ($\beta$-Galactosidase), internal cavities (GroEL), and protruding flexible domains (GDH) are surrounded by residues with high solvent exposure scores and are correspondingly the sites where the largest rearrangements are observed. When there is a coherent, preferred pathway for reorientation, the rearranged structure resolves clearly because then minimal heterogeneity is introduced by dehydration. An incoherent pathway, by contrast, manifests as a region of lower resolution.

We illustrate both regimes with examples from the four proteins. The ESIBD structure of $\beta$-Galactosidase illustrates coherent rearrangement: secondary and tertiary structure are largely preserved, because dehydration-induced subunit reorientation coherently compacts the protein, closing crevices between $\beta$-sheets at the tips of the protein and on the central surface (blue stars in Fig.\,\ref{fig_locres}a,e).\cite{TE2024}

\begin{figure}%[h]
\centering
\includegraphics[width=\linewidth]{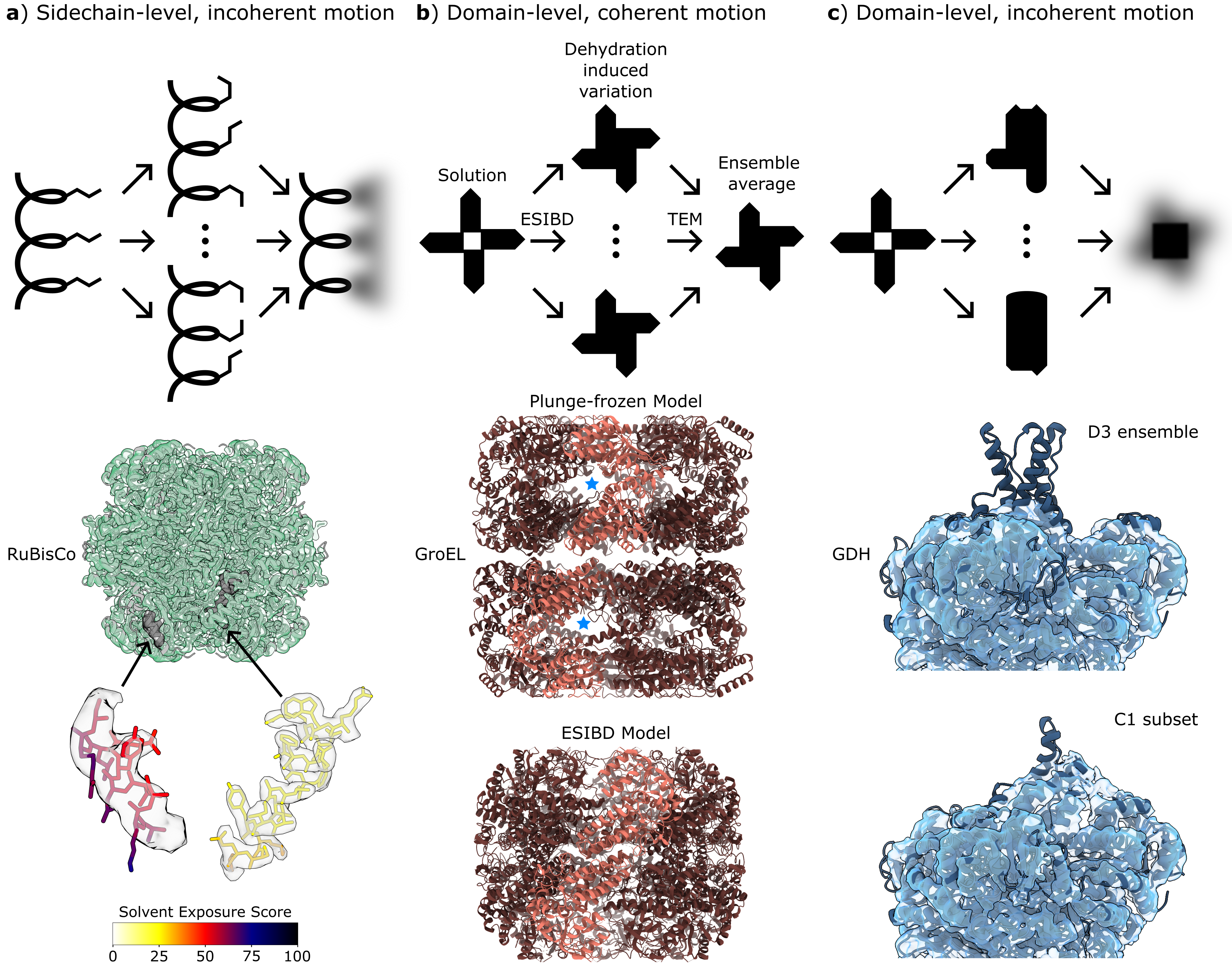}
\caption{\textbf{Dehydration-induced structural changes in ESIBD-prepared proteins.} \textbf{a}) Model of RuBisCo in ESIBD map. Two isolated $\alpha$-helices illustrate the dehydration-induced heterogeneity (lack of density) in highly solvent-exposed sidechains. \textbf{b}) Comparison between a plunge-frozen GroEL model (PDB entry 5W0S) and the ESIBD-prepared model, with two monomers highlighted to demonstrate the twist and compaction between subunits and blue stars labelling crevices which collapse. \textbf{c}) Comparison between consensus D3 map and model and a symmetry-free map and model generated from a subset of particles for GDH, demonstrating the collapse of the three $\alpha$-helices which make up an antenna.}\label{fig_localeffects}
\end{figure}

RuBisCo, by contrast, undergoes only minor compaction (1.5\% across its width) and minimal subunit reorientation. Incoherent rearrangement is nevertheless detectable at the sidechain level: in two representative $\alpha$-helices (Fig.\,\ref{fig_localeffects}a), the helix from the protein interior shows uniformly low solvent exposure scores and all sidechains are well resolved, whereas in a surface helix the most solvent-exposed sidechains are entirely unresolved while the buried ones remain visible.

GroEL exhibits the most pronounced coherent rearrangement: it undergoes a 24\% reduction in height along the C7 axis with almost no change in width perpendicular to it.\cite{SB2025} Comparison of our refined model with a conventional cryoEM model\cite{SR2017} (Fig.\,\ref{fig_localeffects}b; refined using ISOLDE, COOT, and Phenix\cite{TC2018, PE2010, DL2019}) shows that smaller cavities (blue stars in Fig.\,\ref{fig_localeffects}b) have disappeared, and the two heptameric rings twist by 5\textdegree{} around the central C7 axis, enabling tighter compaction (Suppl.\,Movie\,S2). 

Lastly, GDH has two antennae, each comprising three $\alpha$-helices (residues 398–425 in PDB 3JCZ). These antennae are unresolved in the ESIBD map with D3 symmetry imposed. Heterogeneous refinement (Suppl.\,Fig.\,S6) in C1 symmetry yields a subset of particles in which the collapse of one antenna is resolved (Fig.\,\ref{fig_localeffects}c). 

\section*{DISCUSSION}

Across the four protein complexes studied here, dehydration-induced structural change correlates strongly with the local extent of solvent exposure. Both incoherent reorientations, manifesting as resolution loss, and coherent domain movements occur preferentially in regions that were highly solvent-exposed in solution. Highly solvent-exposed regions of proteins have a greater thermodynamic drive to rearrange to form intramolecular interactions to replace protein-solvent interactions lost to dehydration. The strength and spatial imbalance of the interactions that remain after dehydration are augmented by the absence of electrostatic screening that water provides. These factors, new interactions and those of altered strength, combine to drive rearrangements in solvent-exposed regions which we can identify through the simple tool of the solvent exposure score calculated from a PDB structure. Whether such rearrangement actually occurs depends on the local mechanical context: a domain with stabilising neighbours, like the central core of RuBisCo, resists rearrangement, whereas an isolated, weakly-coupled domain, like the GDH antenna helices, rearranges readily even without a coherent pathway.

Mechanistically, these observations are consistent with a kinetic trapping picture. Native ESI retains the fold of the protein, but once in the gas phase as a dehydrated protein ion, the liquid phase structure, which was at thermodynamic equilibrium, is no longer the lowest energy configuration. Due to the absence of solvent, the strength of polar interactions, especially at the protein surface, increases, while hydrophobic interactions are absent.\cite{PW1995} The structure of the protein is not altered immediately upon dehydration, as it can be in a kinetically trapped state. However, small excitations can overcome barriers towards kinetically accessible rearrangements. If these rearrangements follow the same trajectory, they are coherent (Fig.\,\ref{fig_localeffects}b), and we perceive them as domain-level structural changes. Examples of such coherent changes are the closure of crevices and cavities in GroEL and $\beta$-Galactosidase with only a small loss in resolution. But, in particular at the protein surface, these changes can be random, incoherent motions, which we measure as an increase in heterogeneity and hence in a loss in resolution (Fig.\,\ref{fig_localeffects}c).

Beyond accounting for our direct observations, the solvent-exposure framework provides a microscopic account of a long-standing observation in native MS: protein collision cross sections (CCS) measured by ion mobility spectrometry are systematically smaller than predicted from solution structures.\cite{MB2010, KJ2021, CH2011a, ED2009, EM2015, MD2016} The structural changes we resolve directly---crevice closure in $\beta$-Galactosidase, axial compaction and inter-ring twist in GroEL, antenna collapse in GDH, and sidechain disorder in the most solvent-exposed regions of all four proteins---all reduce the effective volume of the protein in the gas phase, accounting for the discrepancy. The solvent exposure score, computable from a solution structure alone, identifies in advance where such compaction will occur, offering a predictive link between IMS measurements and specific structural rearrangements. Notably, the solvent exposure score for GroEL is markedly larger than that for the other proteins and likewise, the measured CCS of GroEL shows the largest deviation from its solution structure.\cite{MB2010, CH2011a, EM2015, MD2016}

Achieving these results required preserving the (kinetically trapped) solution conformation throughout the ESIBD process. To protect the protein structure during ionisation, we employ gentle native nano-electrospray with low flow rates, low emitter potential, and small ions to screen electric fields.\cite{JF2003, LK2013} During ion transport, we employ low potential gradients in regions where collisions with neutral gas molecules occur. Gentle ion-gas collisions are needed to thermalise the protein ion beam, which is essential for deposition at low landing energies without loss of intensity.\cite{PF2023} Low energy ion-surface collisions are required to minimise the energy transferred to the protein upon landing, which is especially efficient on freestanding membrane substrates.\cite{KA2023a} Finally, we deposit the protein ion onto a substrate at cryogenic temperature to suppress thermally activated structural reorganisation of the adsorbed protein.\cite{TE2022a} Throughout, UHV conditions, shielding from warm surfaces, and water-vapour-free transfer minimise contamination. Chemical purity is ensured by m/z-filtering, removing complexes with incorrect stoichiometry and other solution contaminants (Fig.\,\ref{fig_em}). 

The growth of a thin film of vitreous ice around deposited proteins by precisely controlling sample temperature during ice growth, the partial pressure of water, and the duration of ice growth has proven essential to the observation of high-resolution structure. Our instrumentation was specifically designed to implement this capability following methodology established in molecular beam epitaxy. Temperature control at the substrate is essential to control the mobility of the adsorbed water molecules: high mobility at high temperature allows for the ordering of water into a poly-crystalline layer, whereas the absence or very low mobility at low temperature leads to the formation of nanostructures.\cite{JB2005} Only precise and long-time stable temperature control allowed us to find the conditions where diffusion balances with the growth rate such that we obtain amorphous ice growth to embed proteins in a flat, homogenous layer. We speculate that the major effects of the ice layer are improved protection from radiation damage and re-introduction of some polar and van-der-Waals interactions that had been lost upon dehydration. While at low temperature the protein cannot relax its structure at the domain level, the mobility on the level of the individual water molecule---evidenced by the fabrication of a smooth layer---will allow also for small rearrangements, in particular at the protein's surface where dehydration has the greatest effect. As a consequence we observe improvement over ice-free cryoEM imaging in terms of resolution.\cite{TE2022}

Building on the success with soluble proteins, we are extending the ESIBD-cryoEM approach to membrane proteins, where the additional complexity of the surfactant environment must also be controlled to preserve native structure.\cite{LU2022, AO2023, JF2026} Beyond extension to new sample classes, the m/z-filtering capability unique to ESIBD opens applications inaccessible to conventional cryoEM preparation, including the gas-phase enrichment of low-abundance species\cite{SC2017, FS2010} and the structural investigation of specific protein-ligand complexes.\cite{HO2023a, XW2020} Providing a routine-level structural biology tool would require domain-level rearrangements to be controlled, ideally suppressed---either by exploiting adducts and charge reduction to stabilise native structure during dehydration,\cite{JS2007, JG2018} or by reversing rearrangement post-deposition through controlled thermal excitation, as recently demonstrated for laser flash melting of the embedding ice.\cite{SB2025, WC2025}

\section*{METHODS}

\subsubsection*{Solution Preparation}

GDH (G7882) was purchased from Sigma-Aldrich; the lyophilised powder was reconstituted in 200 mM ammonium acetate (pH 6.9) to a final concentration of 20 \textmu M. $\beta$-Galactosidase\cite{TE2024} and GroEL\cite{TE2022} were prepared as previously described. RuBisCo was prepared using established methods, described further in SI.\cite{UF2017}

All proteins were desalted by eluting through two P6 buffer exchange columns (7326222, Bio-Rad), equilibrated with 200 mM ammonium acetate (pH 6.9). They were then diluted in 200 mM ammonium acetate (pH 6.9) to reach the concentration used for native MS: 10 \textmu M ($\beta$-Galactosidase and GDH), 5 mg / mL (RuBisCo), and 5 \textmu M (GroEL). Buffer exchange was always done on the day of deposition.

\subsubsection*{Native MS}

Native MS was performed as previously described.\cite{TE2022}

\subsubsection*{Preparation of ESIBD cryoEM samples}

Mesh size 400 copper TEM grids with 3 nm amorphous carbon on a lacey carbon film (AGS187-4) were purchased from Agar Scientific and gold TEM grids with mesh size 300 and 2 nm amorphous carbon on a R1.2/1.3 holey gold film (C2-A14nAu30-50) were purchased from Quantifoil. 

Grids were clipped into autogrid carriers then plasma cleaned for 30 seconds before deposition. The clipped, cleaned grid was loaded into the cryoshuttle, where it is held in position by springs. The top was then closed, isolating the grid from the atmosphere during transfer. To keep the grid cleaner when ice is already condensed on the stage, the shuttle was loaded partially into the stage before deposition to achieve good thermal contact, but without exposing the grid to atmosphere, as insertion of the room temperature cryoshuttle heats up the stage, sublimating some water. After 5 minutes of temperature equilibration, the cryoshuttle was fully inserted and ready for deposition with the grid exposed.

Deposition methodology is described in the workflow section above. For ice growth, a leak-valve to a reservoir containing liquid water and water vapour at equilibrium (at room temperature) is opened, raising the partial pressure of water in the chamber to a defined value, typically $5\times10^{-5}\,\mathrm{mbar}$. After a defined time, the valve is shut, and excess water vapour is pumped, reducing the partial pressure of water to a negligible amount and stopping ice growth. Transfer into liquid nitrogen using the Aquilos system is performed as previously described.\cite{TE2024}

\subsubsection*{Movie acquisition and processing}

All micrographs were collected using a Thermo Scientific Krios 300 kV cryo-TEM equipped with a BioQuantum energy filter operated at a slit width of 20 eV and a K3 direct electron detector (both Gatan), located at the COSMIC cryoEM facility. Automated data acquisition was controlled using EPU software (Thermo Scientific). All movies were recorded in the tif format, using a range of defocus settings between -1 and -3 μm, an exposure of 40 $e^- /$\r{A}$^2$, and a magnification of 105,000 corresponding to a pixel size of 0.83 Å.

Data were processed using cryoSPARC.\cite{AP2017} After running Patch Motion Corr. and Patch CTF jobs, particles were picked using template picking, based on templates from manual picking, and extracted. After multiple rounds of 2D and 3D classification, final maps were produced using local refinement ($\beta$-Galactosidase, GDH, RuBisCo) or non-uniform refinement (GroEL), based on ab initio initial volumes generated from our data. This was followed by local resolution estimation. Symmetry was imposed in non-uniform refinement and local refinement, but using C1 symmetry resulted in only slightly lower resolution. Figures and movies of the resulting 3D EM density maps were generated using ChimeraX.\cite{EP2021} Further movie acquisition and processing details are specified in figs. S3 to S6 and table S2.

\subsubsection*{Solvent Exposure Score}
 
Using a python script, the solvent exposure score is calculated for each non-hydrogenic atom, $i$, in the plunge-frozen structure of the proteins ($\beta$-Galactosidase: 6CVM; GDH: 3JCZ; RuBisCo: provided by J. Bolla; GroEL: 5W0S). We sum up the $m_jd_{ij}^{-2}$ where $d_{ij}$ is the distance (measured in Ångstrom) between the atom $i$ and all other atoms $j$ (with atomic mass $m_j$) for all atoms within 50 Å which are not covalently bound to atom $i$ ($1.85$ Å $< d_{ij} < 50$ Å). We then subtract the resulting score from a maximal score (404.3) obtained from all proteins considered with water modelled in, and then divide by the maximal score to normalise the final score on a scale from 0 to 100 (though negative values are, in principle, possible). This yields a score which represents solvent-exposed parts with a high value and concealed parts with a low value. This is visualised in ChimeraX by colouring the protein by the calculated scores; when we display the proteins in cartoon view, we averaged the scores of each atom within each residue to choose the displayed score for said residue. The plots and atomic models still display all atoms, without averaged scores. Code is available at \url{https://github.com/lukasaerik/solvent_exposure}

\subsubsection*{Tomogram acquisition and processing}

Tomographic data were acquired with the Titan Krios described above. Tilt series were collected with SerialEM\cite{DM2005} using dose-symmetric acquisition (± 60°) with 3° tilt increments at a magnification of 81000 and pixel size of 1.106 Å/pixel. Image frames (0.24 s exposure/frame) were collected at a dose rate of 11.3 e-/pixel/s at a target defocus of -4 µm. Tomograms were reconstructed using IMOD.\cite{JK1996}

\begin{acknowledgments}

We wish to acknowledge support from Thermo Fisher Scientific who provided the Q Exactive UHMR mass spectrometer and the Aquilos sample transfer system within the framework of a technology alliance partnership. We thank E. Silvester for her help with tomogram acquisition. We wish to acknowledge support from the COSMIC microscope facility, especially from R. Matadeen. 

\subsection*{Funding}

This research was supported by the BBSRC (BB/W017024/1, BB/V019694/1) and the EPSRC (EP/V051474/1). L.T.S. is funded by the Wellcome Trust (218482/Z/19/Z; Wellcome-funded 4 year PhD program in Cellular Structural Biology). Research in the J.R.B. laboratory is supported by the Royal Society through the University Research Fellowship grant (URF/R1/211567). L.E. acknowledges DPhil funding from Vertex Pharmaceuticals.

\subsection*{Author Contributions}

L.E., T.K.E., and S.R. conceived the experiments. L.E., T.K.E., P.F., and S.R. designed and constructed the custom deposition hardware. T.R. and J.R.B. expressed and purified RuBisCo. L.E., T.K.E., and S.B.K. prepared solutions, performed native ESIBD, imaged, and performed data analysis for cryoEM samples. L.T.S. and L.B. collected and built all tomography data. L.E. and M.G. prepared solutions, performed ESIBD, and imaged STM samples. L.E. and S.R. drafted the manuscript. All authors contributed to the interpretation of results and reviewed the manuscript.

\subsection*{Competing Interests}

T.K.E. is an employee of Thermo Fisher Scientific, manufacturer of the Q Exactive UHMR, Aquilos, Arctica, and Krios instruments used in this research.

\subsection*{Data Availability Statement}

cryoEM maps are available in the Electron Microscopy Data Bank (EMDB) under the following accession codes: EMD-57226 ($\beta$-Galactosidase), EMD-57223 (GDH; D3), EMD-57224 (GDH; C1 Subset), EMD-57225 (RuBisCo), and EMD-52626 (GroEL). Models are available in the RCSB Protein Data Bank (PDB) under the following accession codes: 29KD ($\beta$-Galactosidase), 29KB (GDH), 29KC (RuBisCo), and 29KA (GroEL).

\end{acknowledgments}

\clearpage

\setcounter{page}{1} % restart counting 
\setcounter{figure}{0} % restart counting 
\setcounter{table}{0} % restart counting 
\renewcommand{\thefigure}{Fig.~S\arabic{figure}} % add S infront of number
\renewcommand{\thetable}{Table~S\arabic{table}} % add S infront of number
\renewcommand{\theHfigure}{Supplement.\thefigure} % to create unique labels for hyperref
\renewcommand{\theHtable}{Supplement.\thetable} % to create unique labels for hyperref

\begin{center}
{\Large Supplementary Materials for}\\
\textbf{\large High-resolution cryoEM structure determination of soluble proteins after soft-landing ESIBD.}
\end{center}
\noindent Lukas Eriksson \textit{et al.}

\noindent Corresponding Author: stephan.rauschenbach@chem.ox.ac.uk

\subsection*{This PDF file includes}
Supporting text \\
Figures S1 to S6 \\
Tables S1 and S2 \\
Equations S1 to S3 \\
SI references \\

\subsection*{Other Supplementary Materials for this manuscript include the following:}
Movies S1 and S2

\clearpage

\section*{Preparation of RuBisCo}

\textit{Arabidopsis thaliana} seedlings were grown on Murashige and Skoog medium plates (0.65\% agar) to 14d, then macerated in chloroplast isolation buffer (CIB; 0.3 M sorbitol, 20 mM HEPES-KOH pH 8.0, 10 mM NaHCO$_3$, 5 mM MgCl$_2$, 5 mM EDTA, 5 mM EGTA). Macerated cells were filtered through two layers of Miracloth (Merck, Darmstadt) and pelleted for 5 minutes at 1000 x g. Cell pellets were resuspended in CIB buffer and layered over a preprepared 50\% percoll (Cytiva, Uppsala) in CIB gradient. Broken and whole chloroplasts were separated by density gradient centrifugation for 10 minutes at 7800 x g. Whole chloroplasts were washed in HMS buffer (0.3 M sorbitol, 50 mM HEPES-NaOH pH 8.0, 3 mM MgSO$_4$), then lysed for 1 hour in hypotonic lysis buffer (25 mM HEPES-KOH pH 8.0 with protease inhibitor). The stroma of lysed chloroplasts was obtained by laying lysed chloroplasts over the 0.3 M MOPS (0.3 M sucrose, 10 mM MOPS pH 7.8, 4 mM MgCl$_2$) step of a sucrose step gradient and centrifuged at 80000 x g to separate soluble and membrane fractions. Large complexes were concentrated using a 100000 Da molecular weight concentrator.

\section*{Preparation of ESIBD STM Samples}

Single crystals of gold and copper were obtained from MaTeck GmbH, with the (111) and (100) faces polished, respectively. To achieve a clean, atomically-flat surface, 3 cycles of sputtering (Ar, p = 10$^{-5}$ mbar, 1 kV DC electron gun) and annealing (with resistive heater, to 800 K for 5 mins) were used. Samples were then transferred into a UHV vacuum suitcase,\cite{SK2012} removed from the SPM by venting the loadlock, and carried to the ESIBD instrument; the loadlock was pumped to $2\times10^{-8}$\,mbar, and the sample was transferred into the deposition stage.

Three classes of molecules were deposited for proof-of-principle scanning tunneling microscopy (STM) experiments: $\beta$-cyclodextrin, a 1135 Da cyclic oligosaccharide; bovine serum albumin (BSA), a 66 kDa protein; and pUC19, a 1.7 MDa plasmid; representative micrographs are shown in Fig.\,S2. $\beta$-Cyclodextrin (C4767), purchased from Sigma Aldrich, was dissolved into a 50:50 v:v solution of deionised water (MilliQ):HPLC grade methanol to yield a final concentration of 10 \textmu M. Bovine serum albumin (A0281) was purchased from Sigma Aldrich, then 9 mg were dissolved into into 24.75 mL deionised (MilliQ) water, 24.75 mL HPLC grade acetonitrile, and 0.5 mL HPLC grade formic acid. The plasmid pUC19 (SD0061), in a 10 mM Tris-HCl and 1 mM EDTA buffer, was purchased from Thermo Scientific. This was then diluted in a solution of 66\% v/v ACN (in deionised (MilliQ) water) to a final plasmid concentration of 0.3 nM.

Solution was added to gold-coated (with a sputter coater: 108A/SE, Cressington) nano-ESI emitters pulled from borosilicate glass capillaries (30-0042, Harvard Bioscience) using a pipette puller (P-1000, Sutter Instrument). Normal MS setting were used, with lower HCD cell pressure for thermalising the ion beam. First, a mass spectrum without mass selection was obtained, and then mass selection was used to isolate and deposit only the desired species. For $\beta$-cyclodextrin, individual adducts (specifically just the Na$^+$ adduct) were selected for deposition. Sample potential was set to -9 V for a landing energy of 2 eV per charge. Charge accumulated was monitored to measure the amount of ions deposited: 2 pAh for $\beta$-cyclodextrin, 16 pAh for bovine serum albumin, and 10 pAh for pUC19.

After deposition, the sample was transferred into the vacuum suitcase, the loadlock was vented, the suitcase was carried to the SPM, and the loadlock on the SPM was pumped to $2\times10^{-8}$\,mbar. The sample was then transferred into the microscope head, allowed to stabilise in temperature (9.3--9.7\,K). Samples were imaged using scanning tunnelling microscopy in constant current mode with tunnelling currents between 1 and 100 pA and images were processed in Gwyddion.\cite{DN2012}

\section*{Equations}
\begin{equation}
    15\times10^{-12}\,Ah \times \frac{3600\,s}{1\,h} \times \frac{1\,z}{1.602176634\times10^{-19}\,C} \times \frac{1\text{ GDH}}{36\,z} \approx 9.4\times10^{9}\text{ GDH molecules}
\end{equation}
\begin{equation}
    10^{10}\text{ GDH molecules} \times \frac{10^{-5}\,mm  \times 10^{-5}\,mm}{1\text{ GDH molecule}}\,\div\, 4\,mm^2 \approx 25\%\text{ of monolayer coverage}
\end{equation}
\subsection*{Solvent Exposure Score}
\begin{equation}
    S_i = 100 \times (S_\mathrm{MAX} - \sum_j m_j d_{ij}^{-2})
\end{equation}

\begin{figure}[!htb]
  \begin{center}\includegraphics[width=\textwidth]{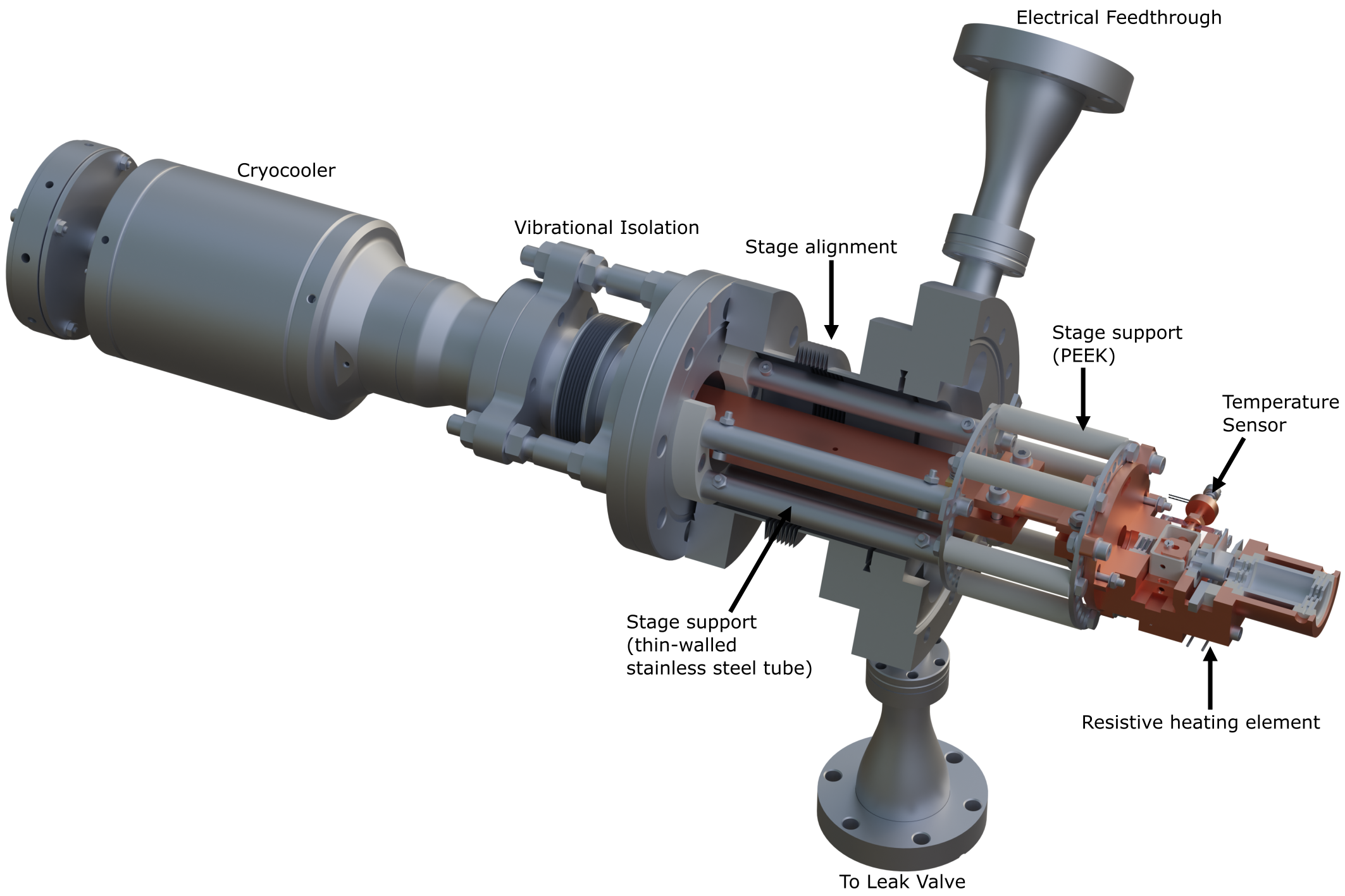}\end{center}
  \caption{\textbf{Deposition stage, support, and cryocooler} shown in half-section view of mounting flange. Key components are labelled.}
  \label{fig_construction}
\end{figure}

\begin{figure}[!htb]
  \begin{center}\includegraphics[width=\textwidth]{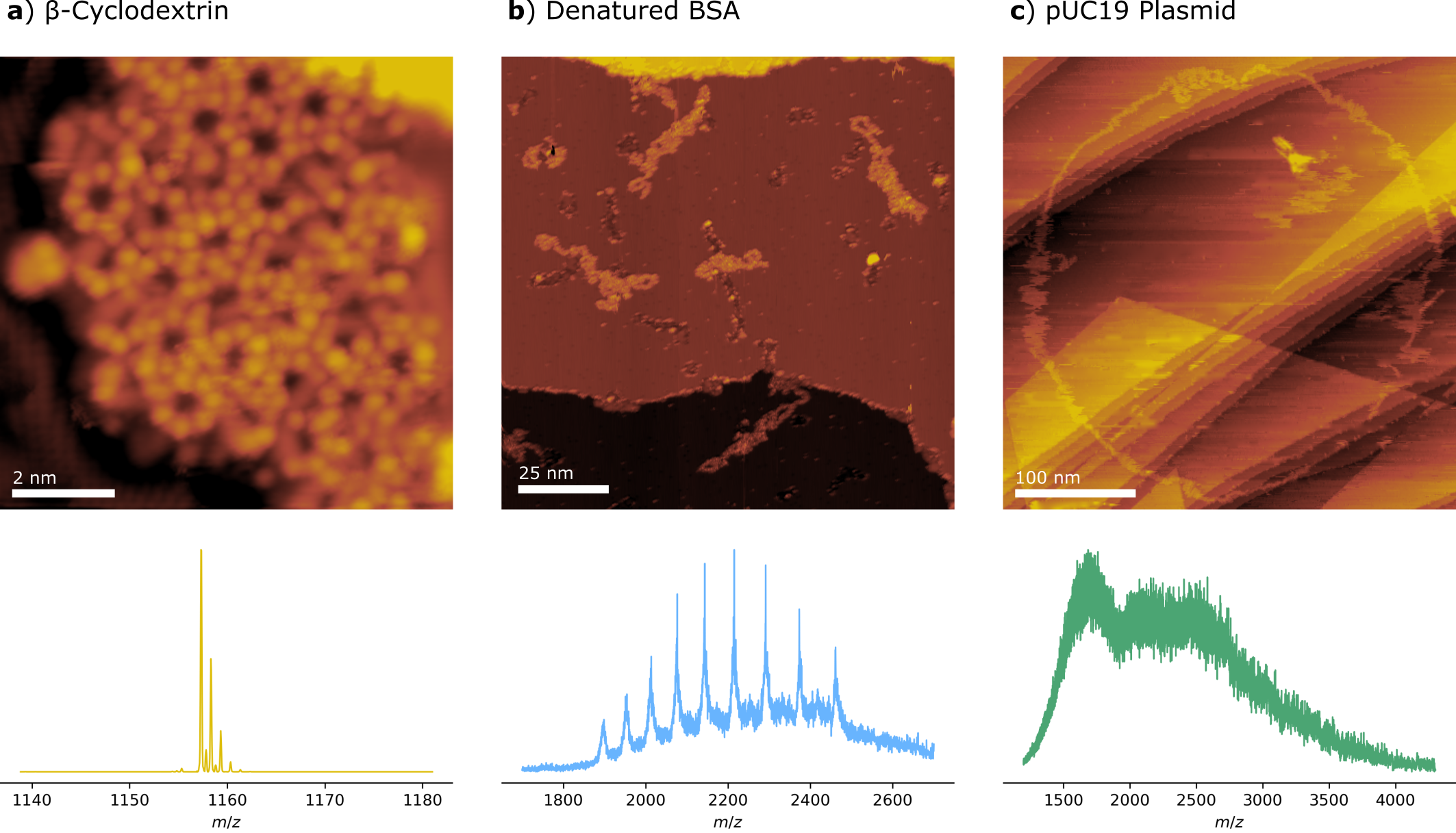}\end{center}
  \caption{\textbf{ESIBD-prepared STM samples.} Representative micrographs and deposited mass spectra for (\textbf{a}) $\beta$-cyclodextrin on Au(111), \textbf{b}) denatured bovine serum albumin on Cu(100), and \textbf{c}) pUC19, a plasmid, on Au(111).}
  \label{fig_stm}
\end{figure}

\begin{table}[!htb]
\caption{\textbf{CryoEM data collection settings for ice micrographs and cryoET.}}
\label{tab_arctica_tomo}
\scriptsize
\begin{tabular}{lll}
\hline
    & Ice Micrographs & CryoET \\
\hline                                                                      
\textbf{Data collection} \\
Microscope                              & Talos Arctica   & Krios G3                        \\
Magnification                           &  	105,000 &	81,800             \\
Voltage (kV)	                        &200        &	300                                  \\
Electron exposure ($e^{-}$/\r{A}$^2$)	& 40        &	2.22 per frame (90.9 total)                        \\
Defocus (\textmu m)	                    & -3        & -4                 \\
Pixel size (\r{A})	                    &1.17       &	1.106                                    \\
\hline
\end{tabular}
\end{table}

\begin{figure}[!htb]
  \begin{center}\includegraphics[width=\textwidth]{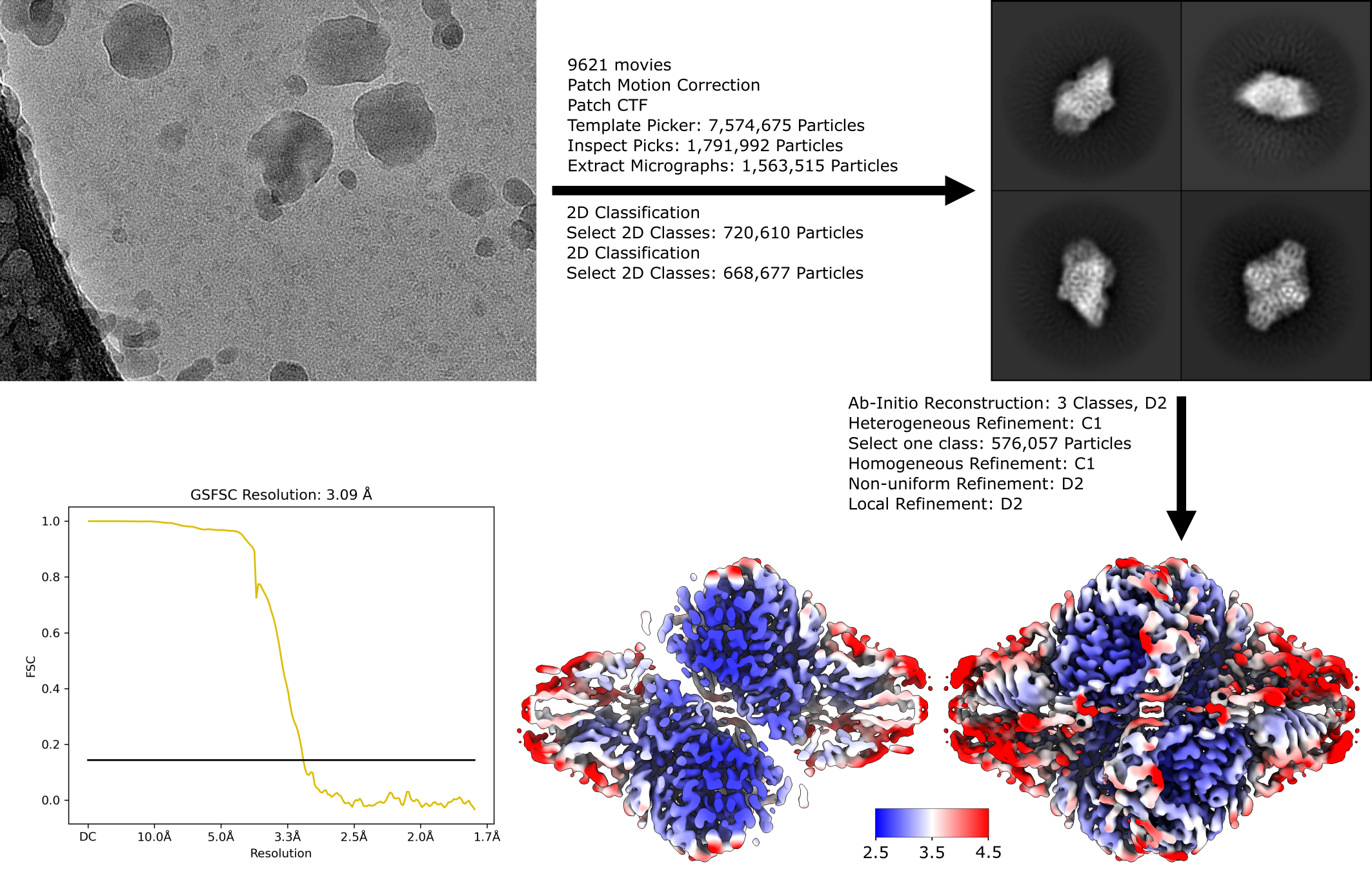}\end{center}
  \caption{\textbf{$\beta$-galactosidase processing pipeline and local resolution map.} \\ }
  \label{fig_bgal}
\end{figure}

\begin{figure}[!htb]
  \begin{center}\includegraphics[width=\textwidth]{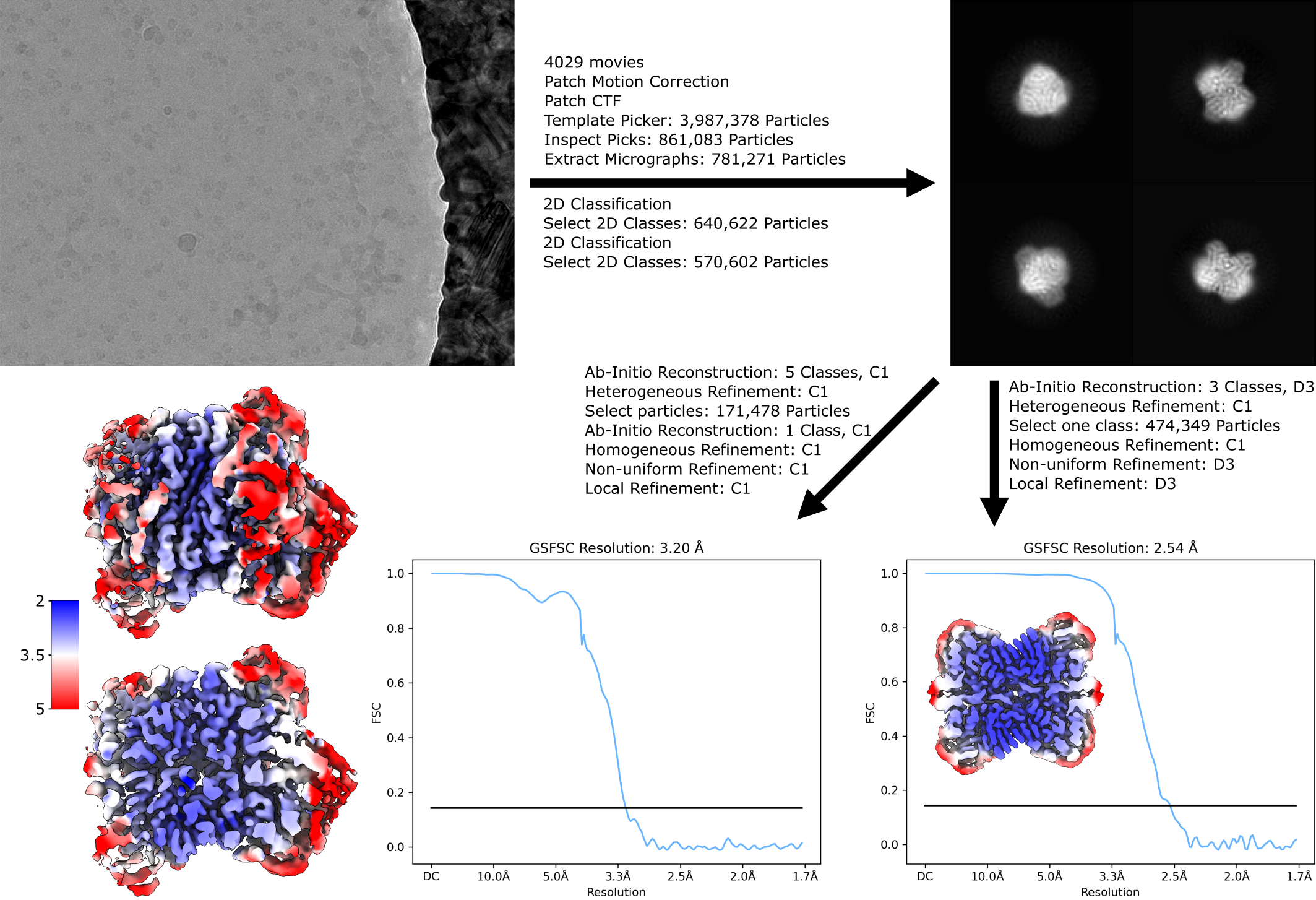}\end{center}
  \caption{\textbf{Glutamate Dehydrogenase (GDH) processing pipeline and local resolution map of C1 subset.} \\ }
  \label{fig_gdh}
\end{figure}

\begin{figure}[!htb]
  \begin{center}\includegraphics[width=\textwidth]{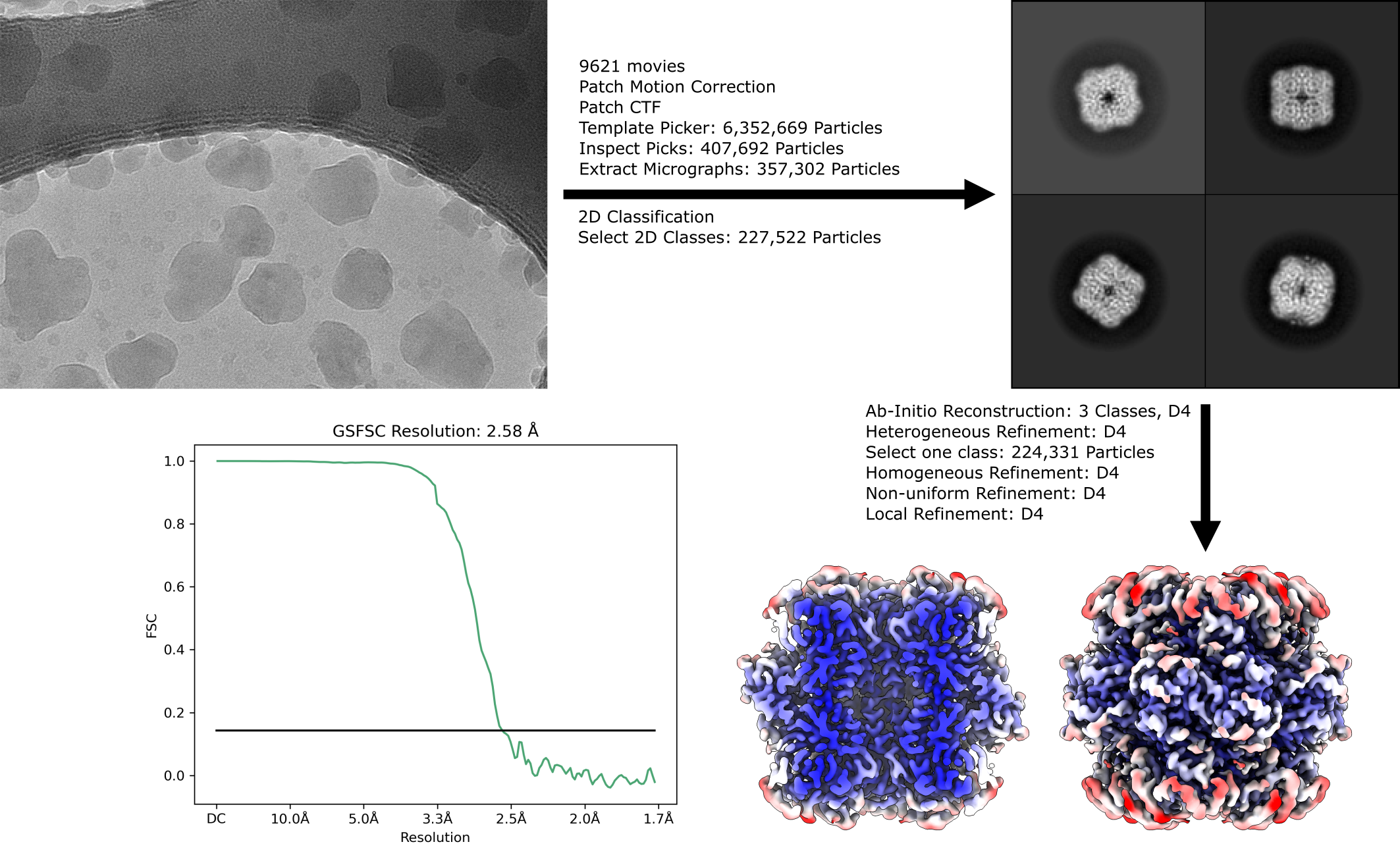}\end{center}
  \caption{\textbf{RuBisCo processing pipeline.} \\ }
  \label{fig_rubisco}
\end{figure}

\begin{figure}[!htb]
  \begin{center}\includegraphics[width=\textwidth]{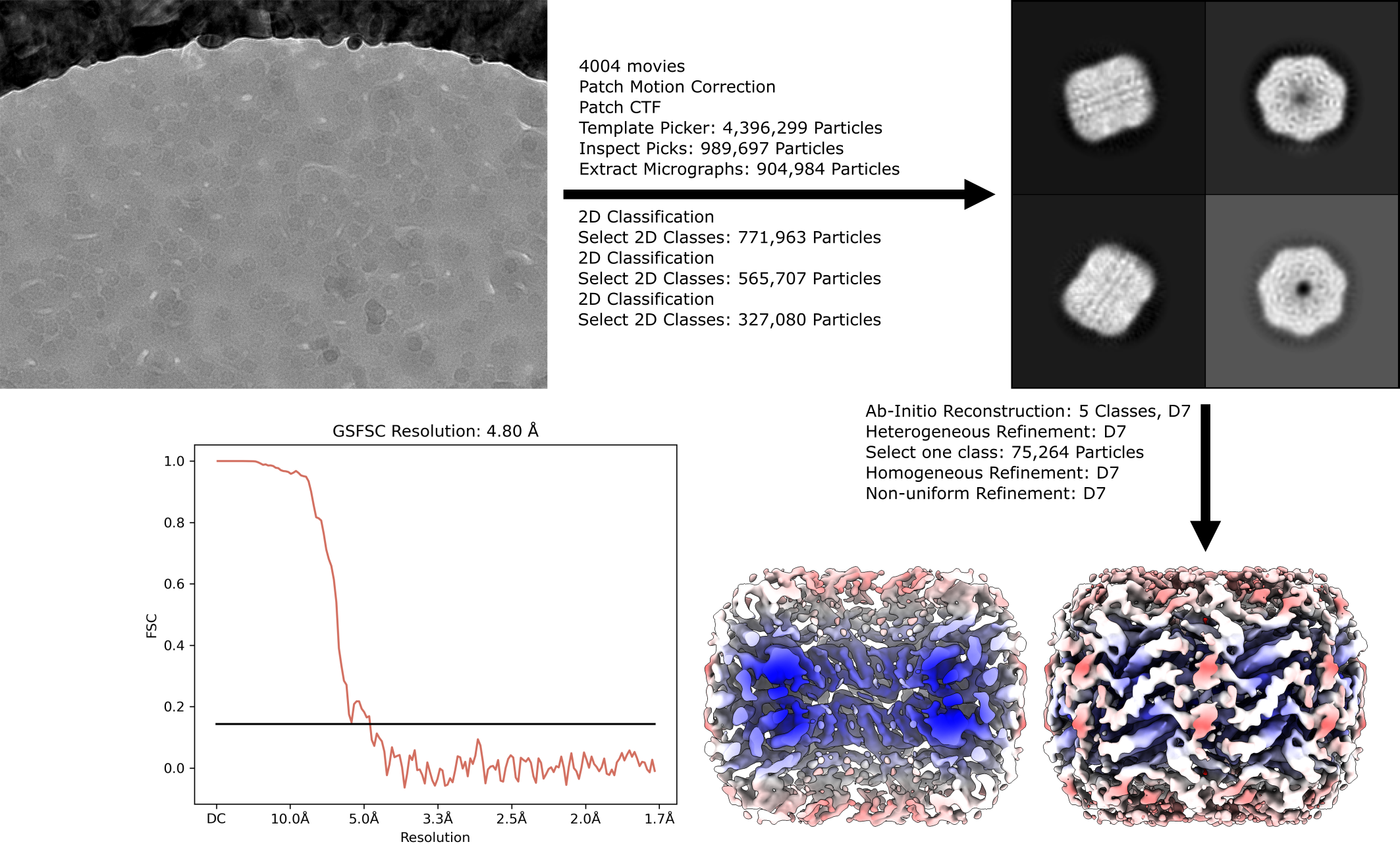}\end{center}
  \caption{\textbf{GroEL processing pipeline.} \\ }
  \label{fig_groel}
\end{figure}

\begin{table}[!htb]
\caption{\textbf{Cryo-EM data collection, refinement, and validation statistics.}}
\label{tab_collection}
\scriptsize
\begin{tabular}{llllll}
\hline
& $\beta$-galactosidase & GDH & GDH (C1)  & RuBisCo    & GroEL \\
& EMD-57226  & EMD-57223 & EMD-57224 & EMD-57225 & EMD-52626 \\
& PDB: 29KD  & PDB: 29KB &           & PDB: 29KC & PDB: 29KA \\
\hline                                                                      
\textbf{Data collection and processing} \\
Microscope                              & Krios G3 & Krios G3 & Krios G3 & Krios G3 & Krios G3 \\
Magnification                           & 105,000 &	105,000 & 105,000 & 105,000 & 105,000 \\
Voltage (kV)	                        & 300 &	300 & 300 & 300 & 300 \\
Electron exposure ($e^{-}$/\r{A}$^2$)	& 40  &	40  & 40  & 40  & 40  \\
Defocus range (\textmu m)	            &-1.5 to -3 & -1 to -2.5& -1 to -2.5&-1.5 to -3 & -1 to -2.5 \\
Pixel size (\r{A})	                    & 0.83 & 0.83 &	0.83 & 0.83 & 0.83 \\
\\
Symmetry imposed                        & D2 & D3 & C1 & D4 & D7 \\
Initial particle images (no.)           & 668,677 & 570,602 & 570,602 & 227,522 & 327,080\\
Final particle images (no.)             & 576,057 & 474,349 & 171,478 & 224,331 & 75,264 \\
Map resolution (\r{A})	                & 3.09	  & 2.54    & 3.20    & 2.58    & 4.80  \\
\hspace{5mm}    FSC threshold	        & 0.143	  & 0.143   & 0.143	  & 0.143   & 0.143 \\
Map resolution range (\r{A}, FSC 0.143) & 1.9 to 4.5 & 1.8 to 5.3 & 2.0 to 6.0 & 1.8 to 3.4 & 3.2 to 6.5 \\
\hspace{5mm} FSC 0.5                    & 2.58 to 5.1 & 2.25 to 5.8 & 2.7 to 6.8 & 2.3 to 4.1 & 4.45 to 8.4 \\
\\
\textbf{Refinement}	                    & &	& N/A &  &  \\
Initial model used (PDB code)           & 6CVM & 3JCZ & & N/A & 5W0S \\
Model resolution (\r{A})                & 3.0/3.1/3.3 & 2.5/2.5/18.5 & & 2.5/2.6/2.7 & 4.6/4.7/24.5 \\
\hspace{5mm}    FSC threshold	        & 0/0.143/0.5 & 0/0.143/0.5  & & 0/0.143/0.5 & 0/0.143/0.5 \\
Map sharpening B factor (\r{A}$^2$)	    & -151.5 & -110.1 & -92.5 & -109.2 & -399.6 \\
Model composition &&&&& \\
\hspace{5mm}   Non-hydrogen atoms       & 22208 & 14622 & & 31896 & 23030 \\
\hspace{5mm}   Protein residues	        & 3520  & 2262  & & 4472  & 4690  \\
\hspace{5mm}   Ligands	                & 0	   & 0     & & 0     &   0   \\
R.m.s. deviations &&&&& \\
\hspace{5mm}   Bond lengths (\r{A})     & 0.006	  & 0.003  & & 0.003 & 0.003 \\
\hspace{5mm}   Bond angles (\textdegree)& 0.772   & 0.561  & & 0.673 & 0.535 \\
Validation &&&&& \\
\hspace{5mm}    MolProbity score        & 2.24 & 1.79 & & 2.04 & 1.32 \\
\hspace{5mm}    Clashscore	            & 5    & 5    & & 6    & 1    \\
\hspace{5mm}    Poor rotamers (\%)      & 3.1  & 2.2  & & 1.8  & N/A  \\
Ramachandran plot &&&&& \\
\hspace{5mm}    Favored (\%)	        & 90  & 96  & & 92  & 90  \\
\hspace{5mm}    Allowed (\%)	        & 10  & 4   & & 8   & 9   \\
\hspace{5mm}    Disallowed (\%)	        & 0   & 0.3 & & 0.3 & 0.5 \\
\hline
\end{tabular}
\end{table}

\clearpage

\noindent\textbf{Movie S1}: Animation of the a TEM grid being loaded in the cryoshuttle and subsequent exposure.

\noindent\textbf{Movie S2}: Morph between solution (5W0S) and ESIBD models of GroEL.

\ifdefined\embedded
\else

\setcounter{figure}{0} % restart counting 
\setcounter{table}{0} % restart counting 
% \captionsetup[figure]{name=} % surpress default prefix
% \captionsetup[table]{name=}
% \renewcommand{\thefigure}{Extended Data Fig.~\arabic{figure}} % add S infront of number
% \renewcommand{\thetable}{Extended Data Table~\arabic{table}} % add S infront of number
\renewcommand{\thefigure}{Fig.~\arabic{figure}} % remove S infront of number
\renewcommand{\thetable}{Table~\arabic{table}} % remove S infront of number

\refstepcounter{figure}\label{fgr:stage}
\refstepcounter{figure}\label{fgr:maps}
\refstepcounter{figure}\label{fgr:atomic}
\refstepcounter{figure}\label{fgr:simulation}

\fi

\clearpage

\bibliography{bib}

\end{document}